\documentclass[twocolumn,superscriptaddress, showpacs, amsmath, amssymb, nofootinbib]{revtex4-1}

\usepackage{xcolor}
\pagecolor{white}
\usepackage{graphicx}
\usepackage{bm}
\usepackage{hyperref}
\usepackage[mathlines]{lineno}
\usepackage{placeins}

\usepackage{tcolorbox}
\tcbuselibrary{listings,breakable}

\newtcblisting{cverbatim}{
      arc=2mm,
      boxrule=0.5pt,
      before skip=10pt,
      after skip=10pt,
      top=4pt,
      bottom=4pt,
      colback=gray!20,
      listing only
}
\begin{document}
\sloppy
\newcommand{\blue}[1]{\textcolor{blue}{#1}}
\newcommand{\red}[1]{\textcolor{red}{#1}}
\newcommand{\hipace}{HiPACE\texttt{++}}

\title{\hipace: a portable, 3D quasi-static Particle-in-Cell code}

\author{S.~Diederichs}
\email{severin.diederichs@desy.de.}
 \affiliation{Deutsches Elektronen-Synchrotron DESY, Notkestraße 85, 22607 Hamburg, Germany}
 \affiliation{Lawrence Berkeley National Laboratory, 1 Cyclotron Rd, Berkeley, California 94720, USA}
\affiliation{University of Hamburg, Institute of Experimental Physics, Luruper Chaussee 149, 22607 Hamburg, Germany}

\author{C.~Benedetti}
\author{A.~Huebl}
\author{R.~Lehe}
\author{A.~Myers}

 \affiliation{Lawrence Berkeley National Laboratory, 1 Cyclotron Rd, Berkeley, California 94720, USA}

\author{A.~Sinn}
 \affiliation{Deutsches Elektronen-Synchrotron DESY, Notkestraße 85, 22607 Hamburg, Germany}

\author{J.-L.~Vay}
\author{W.~Zhang}
 \affiliation{Lawrence Berkeley National Laboratory, 1 Cyclotron Rd, Berkeley, California 94720, USA}

\author{M.~Thévenet}

 \affiliation{Deutsches Elektronen-Synchrotron DESY, Notkestraße 85, 22607 Hamburg, Germany}

\date{\today}

\begin{abstract}
Modeling plasma accelerators is a computationally challenging task and the quasi-static particle-in-cell algorithm is a method of choice in a wide range of situations. In this work, we present the first performance-portable, quasi-static, three-dimensional particle-in-cell code HiPACE\texttt{++}. By decomposing all the computation of a 3D domain in successive 2D transverse operations and choosing appropriate memory management, HiPACE\texttt{++} demonstrates orders-of-magnitude speedups on modern scientific GPUs over CPU-only implementations. The 2D transverse operations are performed on a single GPU, avoiding time-consuming communications. The longitudinal parallelization is done through temporal domain decomposition, enabling near-optimal strong scaling from 1 to 512 GPUs. HiPACE\texttt{++} is a modular, open-source code enabling efficient modeling of plasma accelerators from laptops to state-of-the-art supercomputers.
\end{abstract}

\maketitle

\section{Introduction}

Plasma accelerators \cite{Tajima:1979, Chen:1985} enable the acceleration of charged particles over short distances due to their multi-GeV/m field gradients. Although great progress in terms of beam quality and stability has recently been achieved \cite{Lindstrom:2021,Kirchen:2021}, significant advance is still required to make plasma-accelerator-driven applications feasible. The Particle-in-Cell (PIC) method \cite{Hockney:1981,Birdsall:1985} is a reliable tool to simulate plasma acceleration, and PIC simulations play a major role in understanding, exploring and improving plasma accelerators.

Simulation of a multi-GeV plasma-based accelerator typically requires modeling sub-micron-scale structures propagating over meter-scale distances, hence full electromagnetic PIC simulations require millions of time steps due to the Courant-Friedrichs-Lewy (CFL) condition \cite{Courant:1928}, which makes them unpractical. Several methods were developed to circumvent this limitation and enable larger time steps, including running PIC in a Lorentz-boosted frame~\cite{Vay:2007} or using a quasi-static approximation~\cite{Sprangle:1990a,Mora:1996,Whittum:1997}, both of which have proved performant for modeling of high-energy plasma accelerator stages~\cite{Huang:2006,An:2013,Mehrling:PPCF:2014,Vay:2021}.

Besides algorithmic improvements, further speedup can be accomplished from hardware improvement. Accelerated computing is growing in popularity in the supercomputer landscape~\cite{Top500HPC:2020}, and in particular using GPUs (Graphics Processing Units) as accelerators enabled significant speedup in High-Performance Computing (HPC) applications including PIC~\cite{Bussmann:2013,Myers:2021}. The heterogeneity of processor architectures in HPC makes it difficult to maintain a portable codebase but, following modern HPC practices, this challenge can be efficiently addressed with performance-portability layers~\cite{CarterEdwards:2014,Beckingsale:2019,Zenker:2016}.

In this article, we present the portable, three-dimensional, open-source, quasi-static PIC code HiPACE\texttt{++}\footnote{\url{https://github.com/Hi-PACE/hipace}}~\cite{hipacepp}. HiPACE\texttt{++} is written in C++ and is built on top of the AMReX~\cite{Zhang:2019} framework, which provides field data structure, Message Passing Interface (MPI) communications, and a performance-portability layer. In particular, the quasi-static PIC algorithm was adapted to accelerated computing, and HiPACE\texttt{++} demonstrates orders-of-magnitude speedup over CPU implementations as well as near-optimal scaling up to hundreds of cutting-edge GPUs. These performances enable realistic simulations of $1024\times1024\times1024$ cells for 1000 time steps in less than two minutes on modern GPU-accelerated supercomputers. HiPACE\texttt{++} is a complete rewrite of the legacy C code HiPACE \cite{Mehrling:PPCF:2014}.

The article is organized as follows: Section~\ref{sec:algorithm} summarizes the well-known quasi-static PIC algorithm. The GPU-porting strategy is introduced in Sec.~\ref{sec:gpu}. Correctness of the code is demonstrated in Sec.~\ref{sec:correctness}. Sec.~\ref{sec:performance} presents performance results and a novel parallelization strategy improving scalability on accelerated platforms. Additional code features are highlighted in Sec.~\ref{sec:features}.

\section{The quasi-static particle-in-cell algorithm}\label{sec:algorithm}

In a plasma accelerator, a driver perturbs the plasma electrons (the ions, heavier, remain practically immobile) and drives an electron plasma wave. While the driver can be a laser pulse or a particle beam, we hereafter focus on the case of a particle beam (beam-driven wakefield acceleration) for simplicity, as this is what is currently implemented in HiPACE\texttt{++}. In the driver's wake, a witness beam of charged particles can be accelerated with a high field gradient. In most conditions (with the notable exception of witness beam self-injection), the driver and witness beams evolve on a time scale much longer than the plasma response~\cite{Esarey:2009}. The quasi-static approximation~\cite{Sprangle:1990b} (QSA) treats the beams as rigid when computing the plasma response at a given beam location, hence decoupling the beam and plasma evolutions. Under this approximation, the Maxwell equations take the form of Poisson equations, and the scheme is not subject to a CFL condition. Then, the time step is determined by the smallest betatron period of the beams, making it possible to use time steps orders of magnitude larger than conventional electromagnetic PIC~\cite{Mehrling:PPCF:2014}. The algorithm has two main parts: first, from the distributions of the beams, compute the plasma response (computationally expensive). Second, from the plasma fields, advance the beams by one time step (computationally cheap).

\begin{figure}
	\centering
	\includegraphics[trim={0 0 0 0},clip, width=3.375in]{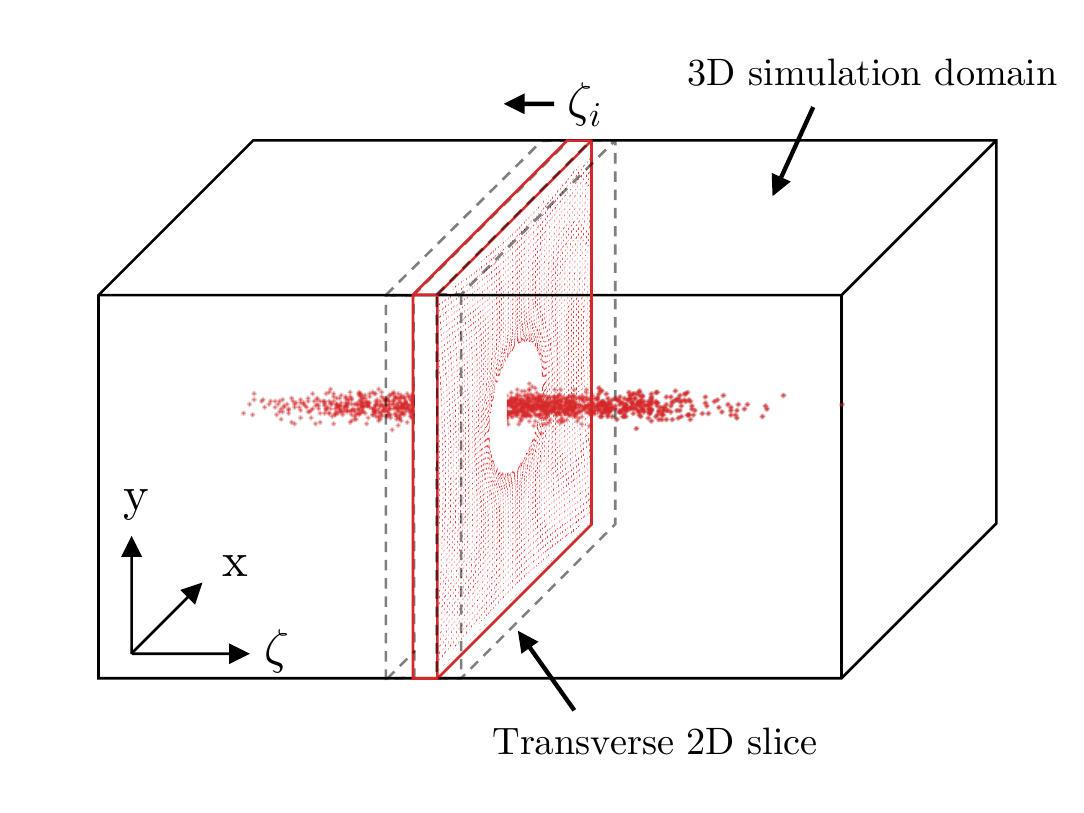}
	\caption{Snapshot of the quasi-static PIC algorithm. The 3D simulation domain is calculated slice-by-slice in a loop over the longitudinal grid points from the head of the box to its tail. Only the beam particles, a 2D slice of plasma particles, and a few 2D slices of fields are required to determine the wake in the 3D simulation domain.}
	\label{fig:qsa_algorithm}
\end{figure}

For a given distribution of the beams, the plasma response is computed in the co-moving frame defined by $\zeta = z - c t$, with $c$ being the speed of light in vacuum (the beams propagate in the $+z$ direction). A slice of unperturbed plasma is initialized ahead of the beams and pushed backwards along the $\zeta$ coordinate. At each longitudinal position, the wakefields are calculated as a 2D problem in the transverse plane. The 3D problem is then solved as $n_\zeta$ 2D transverse problems (called \emph{slices}), with $n_\zeta$ being the number of longitudinal grid points in the simulation domain. Figure~\ref{fig:qsa_algorithm} illustrates the algorithm. Once the fields in the whole 3D domain have been computed, the beams are advanced by one time step.

From Maxwell's equations and the quasi-static approximation, the following field equations can be derived \cite{An:2013}.
The wake potential $\psi = \phi - c\,A_z$, with $\phi$ and $\bm{A}$ being the scalar and vector potential respectively, is obtained from
\begin{equation}
\nabla^2_{\perp} \psi  = -\frac{1}{\epsilon_0}\left(\rho - \frac{1}{c}J_z \right),
\label{eq:psi}
\end{equation}
where $\epsilon_0$ is the vacuum permittivity and $\rho$ and $\bm{J}$ are the total (beams + plasma) charge and current densities, respectively. The transverse fields ${E_x-c\,B_y}$ and ${E_y+c\,B_x}$ are calculated from the transverse derivatives of $\psi$:
\begin{subequations}
\label{eq:exmby}
\begin{align}
    E_x - c\,B_y &= -\frac{\partial}{\partial x} \psi \, , \\
    E_y + c\,B_x &= -\frac{\partial}{\partial y} \psi.
\end{align}
\end{subequations}
The longitudinal field $E_z$ is obtained from
\begin{equation} \label{eq:ez}
\nabla^2_{\perp} E_z = \frac{1}{\epsilon_0 c} \nabla_{\perp} \cdot  J_{\perp} \,.
\end{equation}
The components of the magnetic field are given by
\begin{subequations}
\label{eq:bxby}
\begin{align}
\nabla_{\perp}^2 B_x  &= \mu_0(- \partial_y J_z + \partial_{\zeta} J_y) \,  ,\\
\nabla_{\perp}^2 B_y  &=   \mu_0(\partial_x J_z - \partial_{\zeta} J_x) \, ,
\end{align}
\end{subequations}
and
\begin{equation}
 \nabla^2_{\perp} B_z = \mu_0(\partial_y J_x - \partial_x J_y),
\label{eq:bz}
\end{equation}
where $\mu_0$ denotes the vacuum permeability. All quantities except the longitudinal derivatives $\partial_\zeta J_x$ and $\partial_\zeta J_y$ are directly accessible after the current deposition. These derivatives can be obtained with a predictor-corrector loop~\cite{Huang:2006,Mehrling:PPCF:2014} or by explicit integration~\cite{Wang:2017,Wang:2020}. Both options are available in HiPACE\texttt{++}, hereafter referred to as predictor-corrector or explicit method respectively, and their implementations are described in Sec.~\ref{sec:implementation}. The procedure to calculate fields at slice $\zeta$ given fields at slice $\zeta+\Delta\zeta$ (where $\Delta\zeta$ is the longitudinal cell size) is the standard 2D QSA PIC routine:
\begin{enumerate}
    \item Gather fields and push plasma particles backwards from $\zeta+\Delta\zeta$ to $\zeta$;
    \item Deposit plasma currents and densities;
    \item Deposit beam currents and densities;
    \item Solve equation~\eqref{eq:psi} for $\psi$ to calculate ${E_x-c\,B_y}$ and ${E_y+c\,B_x}$ with equation~\eqref{eq:exmby};
    \item Solve equation~\eqref{eq:ez} for $E_z$;
    \item Solve equation~\eqref{eq:bz} for $B_z$;
    \item Solve equations~\eqref{eq:bxby} for $B_{x/y}$.
\end{enumerate}

\section{Porting quasi-static PIC to GPU}\label{sec:gpu}

\subsection{Porting strategy}

HiPACE\texttt{++} is written considering modern GPU architectures with tens-of-GB global memory, relatively slow transfers between host (CPU) and device (GPU) memories, and fast atomic operations. The algorithm was designed to reduce host-device transfers whenever possible. As illustrated in Fig.~\ref{fig:qsa_algorithm}, the data that needs to be allocated for computation is modest, and only consists in the beam particles, a 2D slice of plasma particles and a 2D slice of grid quantities. In a vast majority of practical cases, these quantities fit in the global memory of a single GPU. This makes it possible to use single-GPU Fast Fourier Transforms (FFTs) that are considerably faster than single- or multi-CPU FFTs, which accounts for a significant fraction of the observed speedup.

In this implementation, the fields on a slice overwrite the previous values, so the fields in the 3D domain are not known at the end of the loop over slices. For this reason, the beam operations (field gather, particle push and current deposition) are also performed per slice within the loop over slices. To that end, beam particles are sorted per slice. Although this results in many small and inefficient kernels (a slice of beam particles contains $\sim$1000 particles for typical simulation parameters), the beam operations take a negligible amount of time overall. Finally, field data in the full simulation domain can be stored for the purpose of diagnostics. In that case, the required data is stacked in host (CPU) memory until the last slice is computed, and then flushed to disk.

\subsection{Implementation}\label{sec:implementation}

Performance-portability is achieved via the AMReX framework~\cite{Zhang:2019}. The most time-consuming functions in the quasi-static PIC method are the field solver, the plasma particle pusher and the plasma current deposition. Particle operations in the PIC method were demonstrated to benefit from GPU computing~\cite{Burau:2010,Myers:2021}, and the same methods as Ref.~\cite{Myers:2021} were applied (in particular, the current deposition relies on fast atomic operations on GPU).

For the \textbf{plasma particle push}, HiPACE\texttt{++} uses a fifth-order Adams-Bashforth particle pusher, as described in \cite{Mehrling:PPCF:2014}. The transverse beam position $x_\perp$, the transverse normalized momentum $u_\perp = \frac{1}{M c} \left( p_x, p_y \right)$, and the normalized plasma wake potential $\Psi_p = \frac{e}{m_ec^2} \psi_p$ of each plasma particle (with $M$ being the mass of the pushed plasma particle, $m_e$ the mass of an electron, and $e$ the elementary charge) are updated as follows:

\begin{align}
\partial_{\zeta}x_\perp ={}& -\frac{u_\perp}{1+\Psi_p } ,  \\[10pt]
\begin{split}\label{eq:1}
\partial_{\zeta}u_\perp ={}&  -\frac{q}{M}\bigg[ \frac{\gamma_p}{1+\Psi_p} \left(
\begin{array}{l}
   E_x - c B_y \\
   E_y + c B_x
\end{array}
\right)
+
\left(
\begin{array}{l}
   \,\,\,\,c B_y \\
   -c B_x
\end{array}
\right) \\ &
+
\frac{cB_z}{1+\Psi_p} \left(
\begin{array}{l}
   \,\,\,\,u_y \\
   -u_x
\end{array}
\right)
\bigg] , 
\end{split}\\[10pt]
\partial_{\zeta}\psi_p ={}&
-\frac{q m_e}{M e} \bigg[ \frac{1}{1+\Psi_p}
\left(
\begin{array}{l}
   u_x \\
   u_y
\end{array}
\right)
\cdot
\left(
\begin{array}{l}
   E_x - c B_y \\
   E_y + c B_x
\end{array}
\right)
- E_z \bigg] , 
\end{align}
with $q$ being the charge of the particle and $\gamma_p$ the Lorentz factor given by
\begin{equation}
\gamma_p = \frac{1 + u_\perp^2 + \left( 1+\Psi_p \right)^2}
                {2\left(1+\Psi_p \right)} .
\end{equation}
The beam particles are advanced by a second-order symplectic integrator. The field gather and particle push are embarrassingly parallel operations well-suited to GPU computing.

Due to the handling per slice, the \textbf{current deposition} is so far limited to zeroth order longitudinally for both plasma and beam particles, while orders 0-3 are available in the transverse direction. On GPU, the current deposition is performed using atomic operations to global device memory.

As can be seen in Eqs.~\eqref{eq:exmby},~\eqref{eq:ez} and~\eqref{eq:bz}, most fields are computed by \textbf{solving a transverse Poisson equation} and applying finite-difference operators. The Poisson equation with Dirichlet boundary conditions is solved by means of fast Poisson solvers~\cite{VanLoan:1992}, which are based on a Discrete Sine Transform (DST) of the first type. The DST is provided by the FFTW~\cite{Frigo:2005} library on CPU and by a custom implementation using FFTs~\cite{Cooley:1970} on GPU. The FFTs on GPU are provided by vendor libraries. The capability to run 2D FFTs on a single GPU instead of parallel FFTs on many CPUs is critical to provide good performance, considering that parallel FFTs require large amount of communications. Special care is needed to compute $B_{x/y}$, because of the longitudinal derivatives $\partial_\zeta J_x$ and $\partial_\zeta J_y$ in Eq.~\eqref{eq:bxby}. The $B_{x/y}$ field solver is usually the most expensive part of the 3D QSA PIC method.

Two options are implemented for the $B_{x/y}$ field solver algorithm. The first option is a \textbf{predictor-corrector field solver}, as implemented in the legacy code HiPACE. The longitudinal derivatives $\partial_\zeta J_{x/y}$ are evaluated on slice $\zeta$ from the previously-computed slice $\zeta+\Delta\zeta$ and slice $\zeta-\Delta\zeta$ still to be computed. An initial guess is made for $B_{x/y}$, with which particles are pushed from slice $\zeta$ to $\zeta-\Delta\zeta$ where their current is deposited. The current on slice $\zeta-\Delta\zeta$ is used to calculate $B_{x/y}$ at $\zeta$, and the procedure is repeated until a convergence criterion is reached or a maximum number of iterations is attained. Each iteration involves all PIC operations for plasma particles as well as several Poisson solves.

The second option for the $B_{x/y}$ field solver is an \textbf{explicit field solver} using analytic integration, as done in Ref.~\cite{Wang:2017,Wang:2020}. A 2D non-homogeneous Helmholtz-like equation must be solved (see equation (19) in Ref.~\cite{Wang:2020}), for which HiPACE\texttt{++} uses the GPU-capable multigrid solver provided by AMReX. The multigrid solver is an expensive operation, but does not required multiple iterations of PIC operations (although an implicit solver happens inside the multigrid solver).

\section{Correctness}
\label{sec:correctness}

The reference setup used throughout this article consists in a typical beam-driven wakefield acceleration simulation containing a driver beam and witness beam with Gaussian distributions with rms sizes $k_p \sigma_{\perp, d} = 0.3$, $k_p \sigma_{\zeta, d} = 1.41$ and $k_p \sigma_{\perp, w} = 0.1$, $k_p \sigma_{\zeta, w} = 0.2$, where $k_p = \omega_p/c$ is the plasma wavenumber, $\omega_p = \sqrt{n_0 e / (m_e \epsilon_0 )}$ is the plasma frequency, and $n_0$ the ambient plasma density (subscripts $d$ and $w$ stand for driver and witness, respectively).

The driver beam is located at the origin and has a peak density of $n_{b, d}/n_0 = 10$. The witness beam is centered around longitudinal position $k_p \zeta_{0, witness} = -5$ and has a peak density of $n_{b, w}/n_0 = 100$. The electron plasma is modeled with 4 particles per cell, and the background ions are assumed to be immobile. The simulation domain in $x$, $y$, and $\zeta$ is, in units of $k_p^{-1}$, ($-8,8$), ($-8,8$), and ($-7,5$) and uses $1024\times1024\times1024$ grid points. All simulation parameters and the used software are listed in the \hyperref[sec:simparam]{Appendix}. All simulations in this section ran on the JUWELS Booster, where each node is equipped with 2 \textsc{AMD EPYC} 7402 processors with 24 cores each and 4 \textsc{NVIDIA A100} GPUs (40\,GB, NVLink3) per node.

\begin{figure}
	\centering
	\includegraphics[trim={0 0 0 0},clip, width=3.375in]{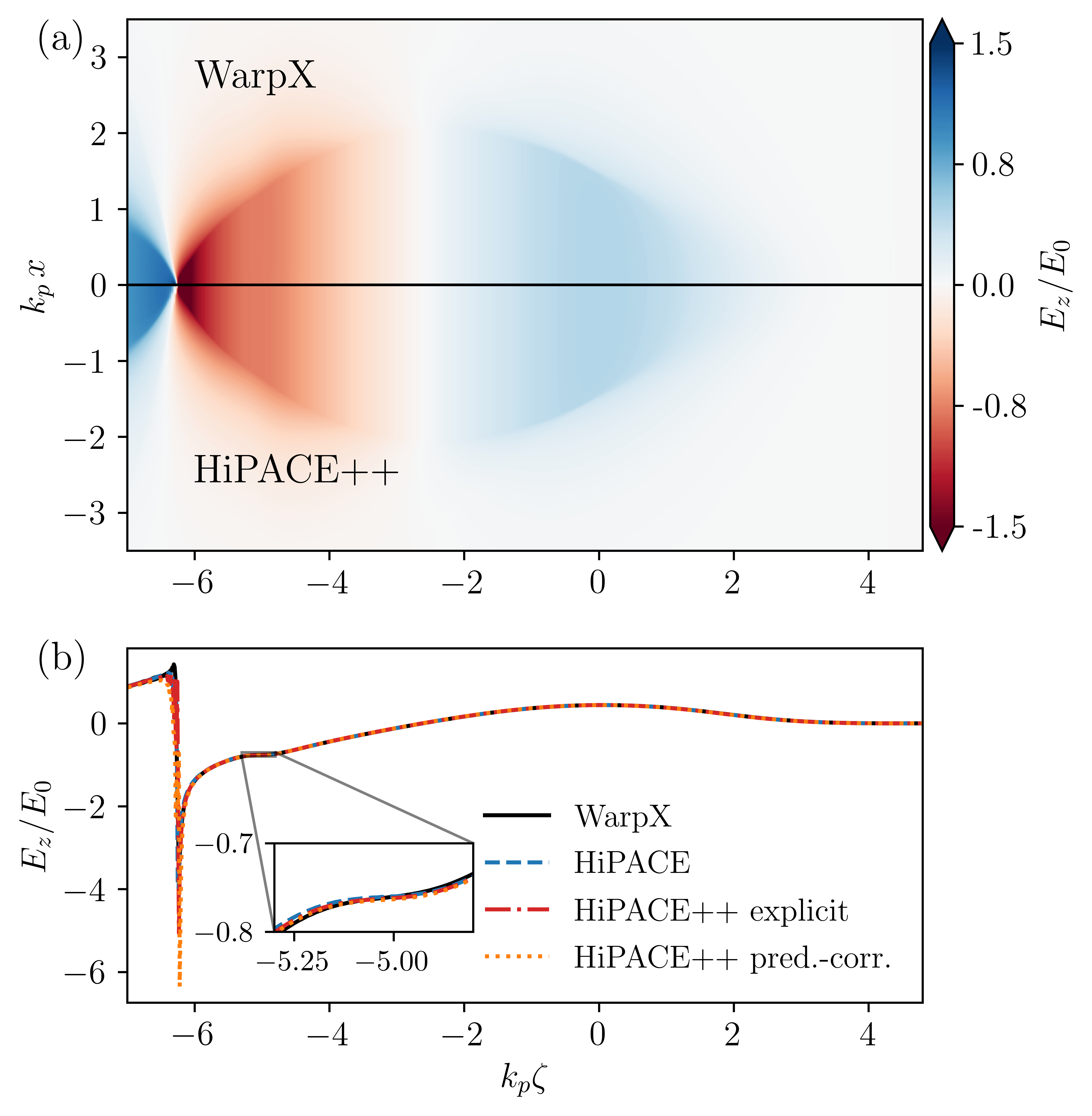}
	\caption{(a) $x$-$\zeta$ snapshot of the electric field in a beam-driven wakefield acceleration simulation using WarpX (top) and HiPACE\texttt{++} (bottom). (b) Lineout of the accelerating field from WarpX, HiPACE, and HiPACE\texttt{++}. The inset shows a zoom on the witness beam region, where flattening of the accelerating field due to beam loading is visible. Both the predictor-corrector (pred.-corr.) and the explicit field solvers are shown.} \label{fig:Ez_comp}
\end{figure}

Figure~\ref{fig:Ez_comp} shows a comparison between HiPACE\texttt{++}, the legacy code HiPACE, and the full GPU-capable 3D electromagnetic PIC code WarpX~\cite{Vay:2021}. The accelerating field $E_z/E_0$, where $E_0 = c m_e \omega_p / e$ is the cold non-relativistic wave breaking limit, shows excellent agreement between these three codes. For HiPACE\texttt{++}, both the predictor-corrector and the explicit solver were used, and both demonstrate again very good agreement, as shown in Figure~\ref{fig:Ez_comp}(b). For WarpX, the rigid beams propagated in a uniform plasma long enough for the wake to reach a steady state. Minor differences in the witness beam region and in the spike at the back of the bubble can be attributed to different physical models, numerical methods, and initialization.

\section{Performance and parallelization}\label{sec:performance}

\subsection{Single GPU performance}

As discussed in Sec.~\ref{sec:gpu} and illustrated in Fig.~\ref{fig:qsa_algorithm}, the amount of data that must be allocated for a 3D domain is relatively modest and consists of beam particles and 2D slices of plasma particles and fields on the grid. Due to the relatively small size of beam data, the amount of data virtually depends only on the transverse number of cells. For example, the total allocated data on the global memory of a NVIDIA A100 GPU with the explicit solver (respectively predictor-corrector method) with 2 million beam particles (accounting for $\sim 230\,\mathrm{MB}$) and 1 particle per cell for the plasma electrons is 2.0 GB (resp. 2.0 GB) for a problem of $128\times128$ cells transversely, 2.9 GB (resp. 2.6 GB) for a $1024\times1024$ problem and 19.2 GB (resp. 12.9 GB) for a $4096\times4096$ problem size (for details on all simulation parameters see the \hyperref[sec:simparam]{Appendix}). Therefore, most practical problems fit on a single GPU, and the performance of HiPACE\texttt{++} on a single NVIDIA A100 GPU is detailed below.

\begin{figure}
	\centering
	\includegraphics[trim={0 0 0 0},clip, width=3.375in]{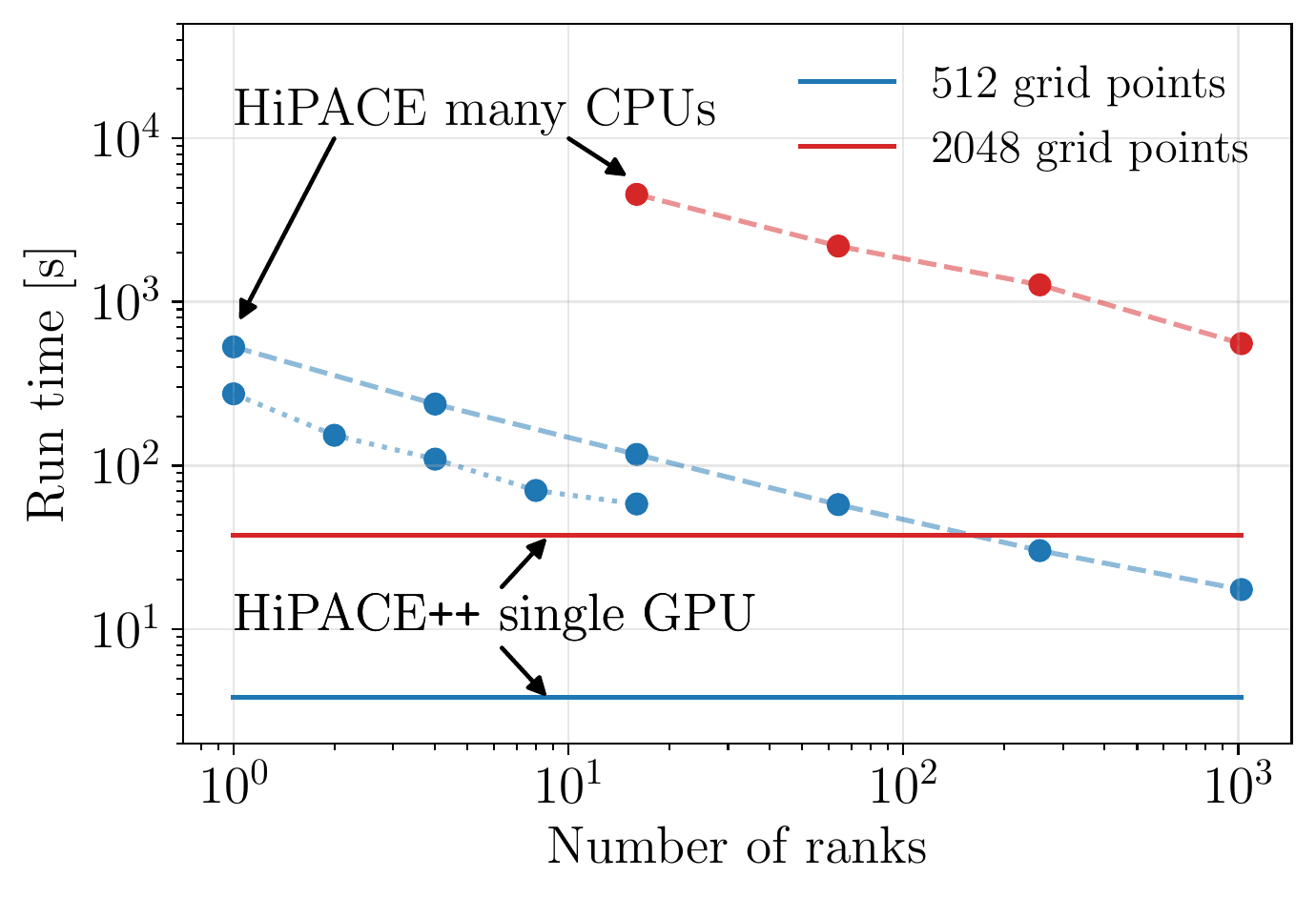}
	\caption{Performance comparison between single GPU and MPI-parallel many-CPU, for the same setup as Sec.~\ref{sec:correctness} for a single time step with medium ($512\times512\times1024$ cells, blue lines) and high ($2048\times2048\times1024$ cells, red lines) resolutions, with predictor-corrector field solver) on the JUWELS Booster. Simulations on CPU used HiPACE (MPI-parallel, dashed lines) and HiPACE\texttt{++} (OpenMP-parallel, dotted line). Simulations on GPU used HiPACE\texttt{++}. The high resolution run with HiPACE did not fit on less than 16 nodes on CPU.}
	\label{FIG:singleGPUvsCPUs}
\end{figure}

The benefits of fitting the problem on a single GPU is clearly demonstrated in Fig.~\ref{FIG:singleGPUvsCPUs}. Typical CPU implementations of the 3D QSA PIC method~\cite{Mehrling:PPCF:2014,Wang:2020,An:2013} accelerate the calculation by decomposing the domain transversely, resulting in large amounts of communications (in particular in the Poisson solver) that dominate the runtime and cause non-ideal scaling. The CPU runs used only the 48 CPU cores on the nodes of the JUWELS Booster. The GPU runs also used the 4 GPUs. On GPU, the simulations at medium and high resolutions take $3.8$\,sec and $37.6$\,sec and cost $2.6 \times 10^{-4}$~node-hours and $2.6 \times 10^{-3}$~node-hours, respectively. For the same simulations using 1024 cores on CPU, HiPACE requires $17.5\,$sec and $556.1\,$sec for a cost of $0.10$~node-hours and $3.3$~node-hours. At medium resolution, the run on 1 (1024) CPU cores was $138\times$ ($4.5\times$) slower and cost $11.5\times$ ($388 \times$) more node-hours than on 1 GPU. At high resolution, the run on 16 (1024) CPU cores was $120\times$ ($14.8\times$) slower and cost $161\times$ ($1262\times$) more node-hours than on 1 GPU. The number of node-hours was calculated as [number of CPU cores]/48 for CPU runs, and [number of GPUs]/4 for GPU runs, as each node has 48 CPU cores and 4 GPUs.

\begin{figure}
	\centering
	\includegraphics[trim={0 0 0 0},clip, width=3.375in]{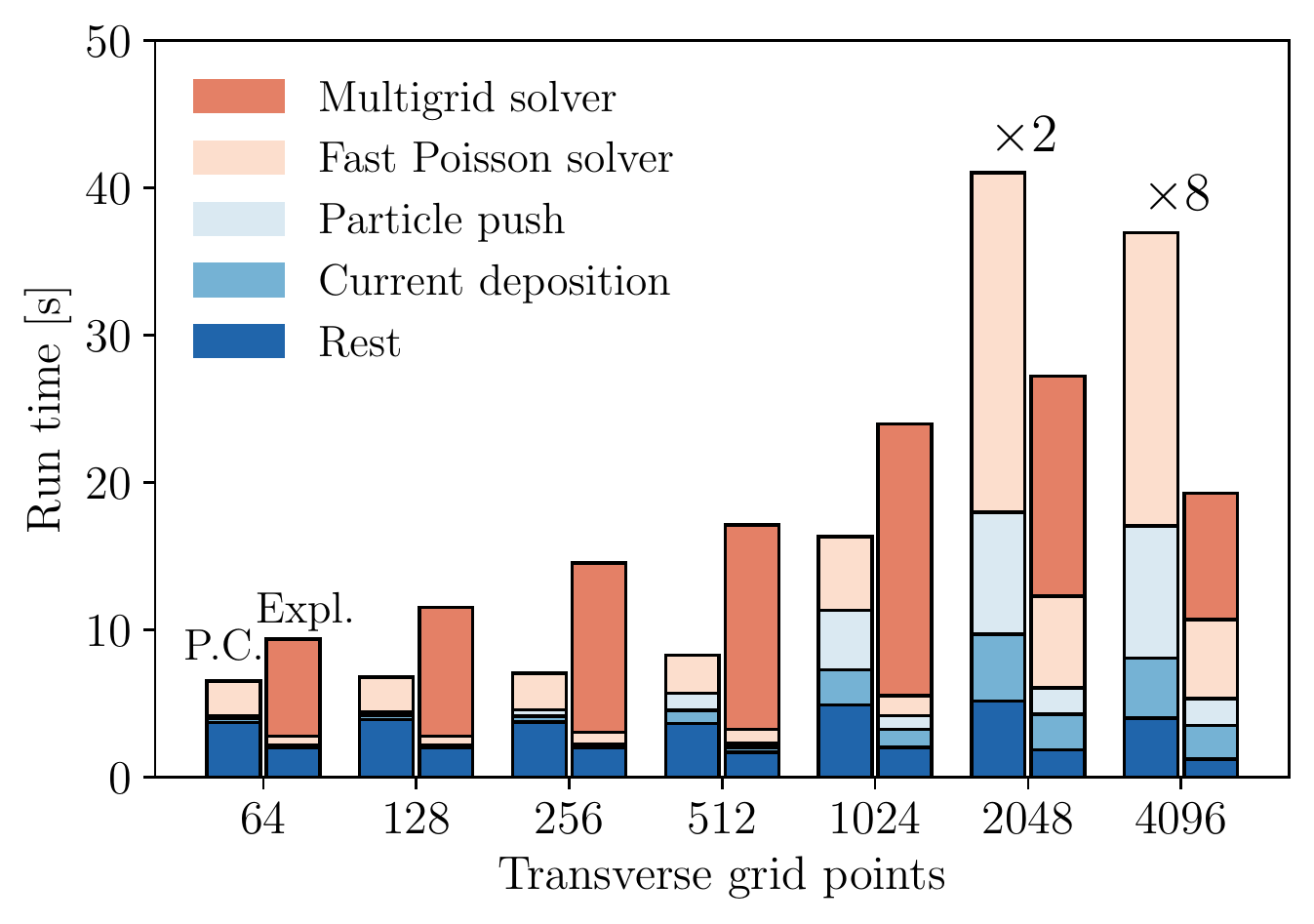}
	\caption{Runtime for different transverse resolutions on NVIDIA A100 GPUs. Left bars: using the predictor-corrector loop. Right bars: using the explicit field solver. The runtimes of $2048\times2048$ and $4096\times4096$ transverse grid points were divided by a factor of 2 and 8, respectively, to improve readability of the figure. 
	\label{FIG:transversescaling}}
\end{figure}

For further insight into the performance of HiPACE\texttt{++}, we ran the reference setup presented in Sec.~\ref{sec:correctness} with increasing transverse resolution, keeping all other parameters constant (for more details see the \hyperref[sec:simparam]{Appendix}). This scan uses 1024 longitudinal grid points, and performance data is given for both the predictor-corrector loop and the explicit field solver. The predictor-corrector loop used up to 5 iterations, so the runtime would be comparable between the two solvers.

The most time-consuming function of the two solvers are shown in Fig.~\ref{FIG:transversescaling}. In both cases a vast majority of the time is spent in solving for $B_{x/y}$. While both the fast Poisson solver and particle operations dominate the predictor-corrector solver at different resolutions, the multigrid solver is always the most expensive operation for the explicit solver. As a reminder, each iteration in the predictor-corrector loop involves all PIC operations for the plasma particles (field gather, particle push, current deposition and field solve) repeated up to 5 times per slice. Note that this study is not a comparison of the two field solvers, as they have different convergence properties, but rather a performance analysis of each solver separately.

Although CPU-only computing is not the main target of HiPACE\texttt{++}, shared-memory parallelization is implemented to enable transverse parallelization when running on CPU only, using OpenMP. In that case, tiling is implemented for plasma particle operations (field gather, particle push and current deposition), and the threaded version of FFTW can be called. As shown by the dotted line in Fig.~\ref{FIG:transversescaling}, the transverse OpenMP parallelization of HiPACE\texttt{++} gives a similar scaling as the pure MPI transverse parallelization of the legacy code HiPACE up to 16 threads (running on 16 cores of the 24-core JUWELS Booster CPUs).

\subsection{Longitudinal parallelization via temporal domain decomposition}
\label{sec:parallelization}

As presented in Sec.~\ref{sec:algorithm}, the computation of the fields in the full domain at a given time step relies on a loop from the headmost slice to the tail, and consequently cannot be parallelized longitudinally by standard domain decomposition. When computing multiple time steps, longitudinal parallelization can be achieved via pipelining algorithms~\cite{Feng:2009,Sosedkin:2016}, which were first realized in the form of a spatial decomposition~\cite{Feng:2009}. Because of the combination of (i) faster computation of each slice and (ii) using a single rank per slice, the standard spatial decomposition demonstrates poor scaling with our GPU implementation. We hereafter present a temporal domain decomposition, more suitable to the GPU implementation in HiPACE\texttt{++}. This implementation has some similarities with the streaming pipeline presented in Ref.~\cite{Sosedkin:2016}. Both pipelines are summarized below, assuming the  problem is decomposed longitudinally in as many sub-domains (\emph{boxes}) as the number of ranks $n_{ranks}$, and runs for \verb!nt! time steps.

\begin{figure}
	\centering
	\includegraphics[trim={0 0 0 0},clip, width=3.375in]{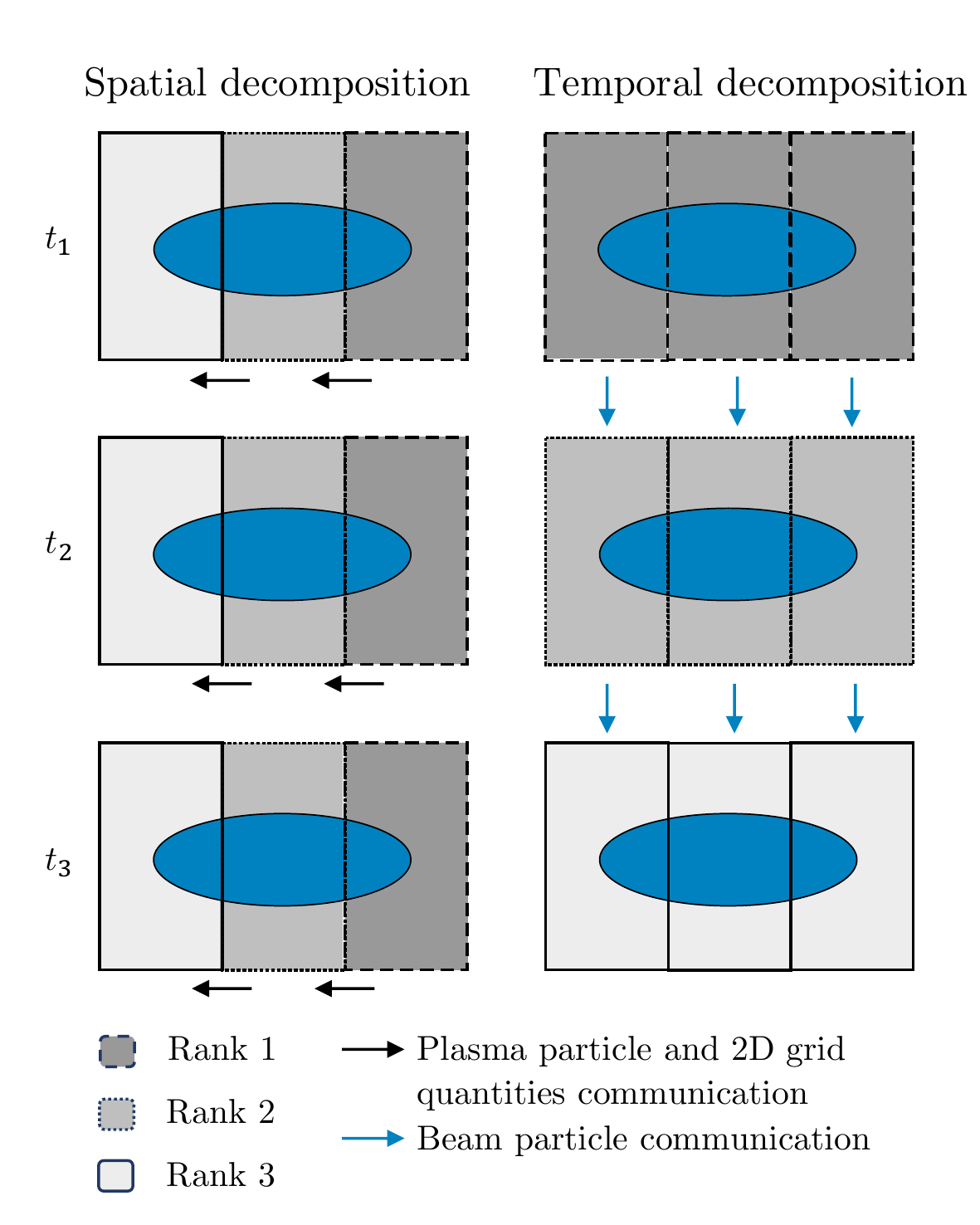}
	\caption{Left: spatial domain decomposition. Each rank calculates a fixed sub-domain for all time steps. Plasma particles and 2D field slices need to be communicated. Right: temporal domain decomposition. Each rank calculates the full domain for a sub-set of time steps. The beam particles of a sub-domain need to be communicated.}
\label{FIG:new_pipeline_scheme}
\end{figure}

In the spatial decomposition, each rank gets assigned one sub-domain, which it consecutively calculates for every time step. The algorithm (in pseudo-code) reads:
\begin{cverbatim}
# Rank r computes box b for all time steps
for t in 0:nt-1:
  Receive last slice from box b+1 at time t
  Compute box b at time t
  Send last slice to box b-1 at time t
\end{cverbatim}
where a slice consists in field data and plasma particle data. This pipeline is represented on the left of Fig.~\ref{FIG:new_pipeline_scheme}. The number of scalars communicated per time step and per rank reads $N_s = n_xn_y(S_{cell} + n_{ppc}S_{plasma})$ where $n_{x}$ ($n_y$) is the number of cells in the transverse direction $x$ ($y$), $S_{cell}$ is the number of scalars communicated per cell (in HiPACE\texttt{++}, $S_{cell}=6$ for $J_x$ and $J_y$ of the previous slice, and $B_{x/y}$ of the two previous slices) and $S_{plasma}$ is the number of scalars communicated per plasma particle (in HiPACE\texttt{++}, $S_{plasma}=35$ due to fifth-order Adams-Bashforth pusher). Here, $n_{ppc}$ is the number of plasma particles per cell. Each rank communicates a full slice, so the amount of data communicated does not scale with $n_{ranks}$.

In the temporal decomposition, each rank computes the full domain for the subset of time steps $t$ for which ${t \equiv r \pmod{nt}}$ (where $r$ is the current rank). The algorithm (in pseudo-code) reads:
\begin{cverbatim}
# Rank r computes all boxes for time step t
for b in nb-1:0:
  Receive beam from box b at time t-1
  Compute box b at time t
  Send beam to box b at time t+1
\end{cverbatim}
This pipeline is represented on the right of Fig.~\ref{FIG:new_pipeline_scheme}. The number of scalars communicated per time step and per rank reads $N_t = n_{beam,r}\times S_{beam}$ where $n_{beam,r}$ and $S_{beam}$ denote the number of beam particles on that rank and number of scalars communicated per beam particle, respectively. As can be seen in Fig.~\ref{FIG:new_pipeline_scheme}, $n_{beam,r}$ scales with the number of ranks (\textit{i.e.}, the number of sub-domains), so this pipeline should perform better for strong scaling.

As an example, let us consider a typical problem with $n_x = n_y = 1024$, $n_{ppc} = 1$, and $n_{beam,total}=2\times 10^6$. Even in the most favorable case (exchanging as few scalars as possible), $S_{plasma}=7$ for position, momentum and particle weight (HiPACE\texttt{++} uses 35), $S_{beam}=7$ and $S_{cell}=6$ to calculate the initial guess, the amount of data (assuming IEEE 754 \verb!double! precision) exchanged per rank and per time step is $110\,\mathrm{MB}$ for the spatial decomposition. Although this is usually not the case, we assume a load-balanced beam particle distribution across ranks for simplicity, so that $n_{beam,r} = n_{beam,total}/n_{ranks}$. The temporal decomposition exchanges roughly $110\,\mathrm{MB}/n_{ranks}$ per rank and per time step. For $n_{ranks}=256$ the temporal decomposition exchanges roughly two orders of magnitude less data than the spatial decomposition, and is hence expected to show better scalability.

The performance of the temporal decomposition pipeline is assessed via a strong scaling of two different setups. One setup is the reference simulation setup from Sec.~\ref{sec:correctness} with $n_{steps}=1000$ time steps and the other uses $2048\times2048\times 2048$ cells and 2048 time steps (for more details see the \hyperref[sec:simparam]{Appendix}). The efficiency $\eta$ is given by $\eta(n_{ranks}) = t(1)/[n_{ranks} t(n_{ranks})]$ where $t(i)$ is the run time on $i$ ranks. Due to the filling and emptying of the pipeline, the ideal efficiency for both pipelines is not identically 1 but rather given by
\begin{equation}
    \eta_{ideal}(n_{ranks}) = \left( 1+ \frac{n_{ranks} - 1}{n_{steps}}\right)^{-1}.
\end{equation}
An efficiency of $1$ is obtained in the limit of $n_{steps} \gg n_{ranks}$.
\begin{figure}
	\centering
	\includegraphics[trim={0 0 0 0},clip, width=3.375in]{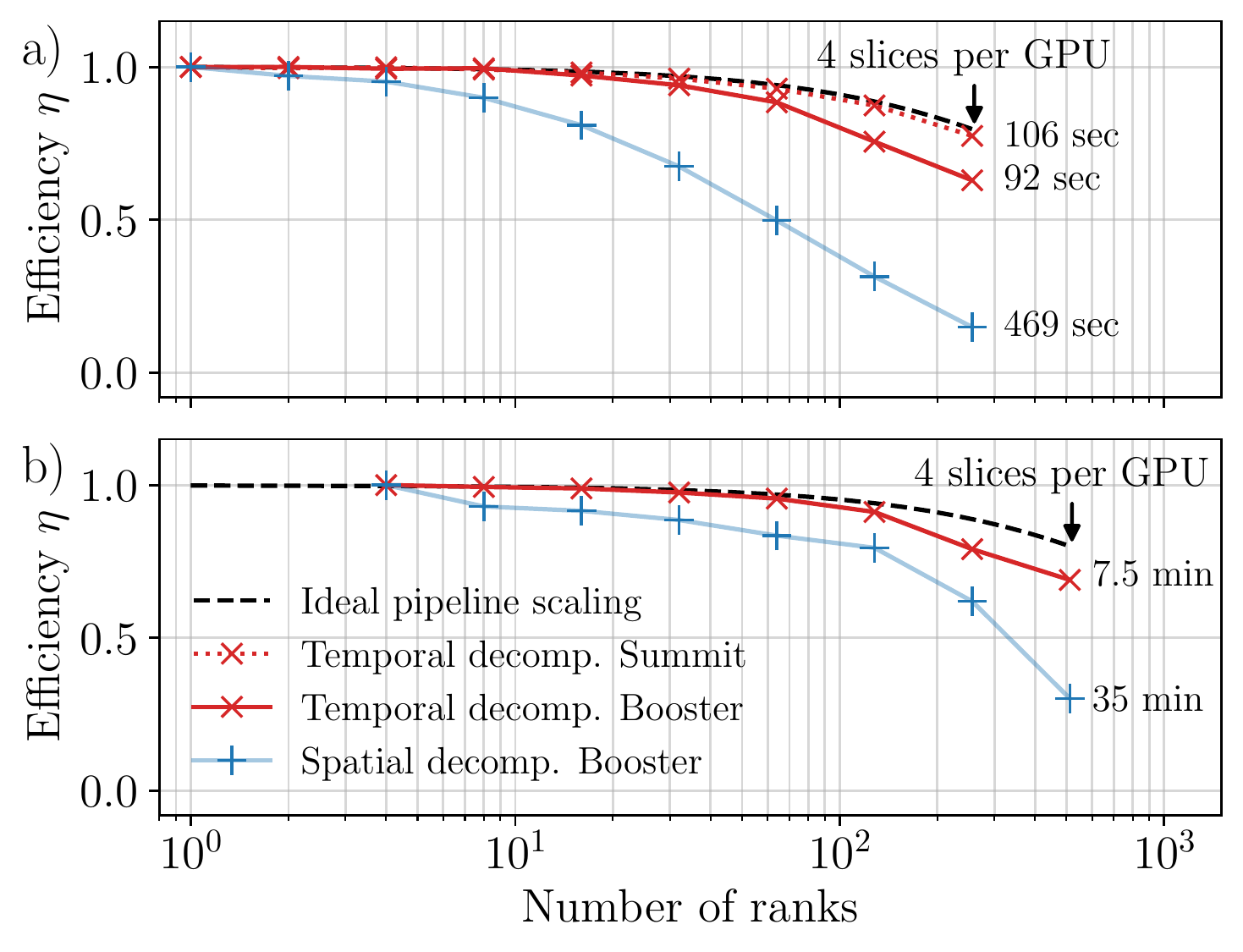}
	\caption{Strong scaling for two different problems with a) $1024\times1024\times1024$ cells with $4$ plasma particles per cell for 1000 time steps and b) $2048\times 2048 \times 2048$ cells with $1$ plasma particle per cell for 2048 time steps. Both settings used two beams with $10^6$ beam particles each. The final run time is given for the maximum number of ranks used. In b) the scaling starts at $4$ GPUs due to time limit restriction on the supercomputer. The problems are parallelized in the longitudinal direction only.}
\label{FIG:strong_scaling}
\end{figure}
The results are shown in Fig.~\ref{FIG:strong_scaling}. The temporal domain decomposition (red lines) shows an efficiency close to the ideal pipeline scaling (black dashed line). The spatial decomposition (blue lines) suffers from efficiency degradation above 8 ranks. Both scalings were performed on the JUWELS Booster and the reference setup was also run on Summit (red dotted line), which is equipped with 6 NVIDIA V100 GPUs per node. The maximum number of ranks is chosen so that only 4 slices remain per sub-domain, which was the case at 256 ranks ($=256$ GPUs) for the reference setup and 512 ranks for the higher-resolution case. The absolute run time between the temporal and spatial decompositions also vary due to significant performance enhancements not related to the parallelization in the runs with the temporal decomposition.

\section{Software practice and additional features}\label{sec:features}
HiPACE\texttt{++} is a versatile open-source, 3D, quasi-static PIC software with an object-oriented design to invite the integration of new numerical methods or physics packages. HiPACE\texttt{++} uses the cross-platform build system CMake and can be installed, as well as its dependencies, with software package managers, such as Spack~\cite{Gamblin:2015}.

HiPACE\texttt{++} complies with the openPMD standard~\cite{Huebl:2015} and uses the openPMD-api~\cite{Huebl:2018} for I/O, allowing for interoperability and simple benchmarking with other codes. Both HDF5~\cite{hdf5} and ADIOS2~\cite{Godoy:2020} file formats are supported (a feature inherited from the openPMD-api), and the capability to read an external beam from file at the openPMD format is available.

Two unit systems are available in HiPACE\texttt{++}, SI units and normalized units, available as a runtime parameter. In normalized units, all lengths are re-scaled to the plasma skin depth $k_p^{-1}$, the fields to cold, non-relativistic wave breaking limit $E_0$, and all densities to the background plasma density $n_0$. 

\begin{figure}
	\centering
	\includegraphics[trim={0 0 0 0},clip, width=3.375in]{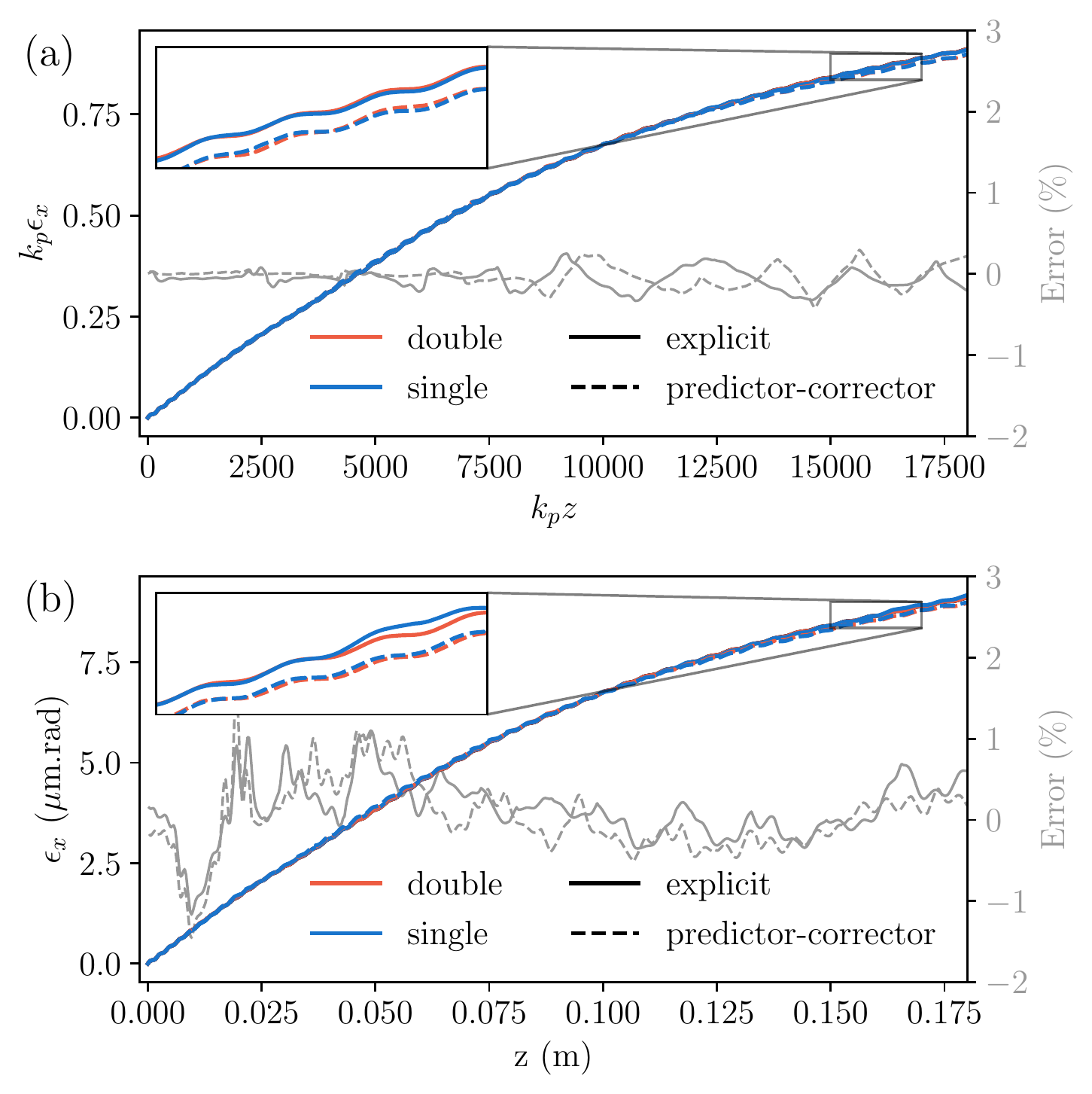}
	\caption{(a) Evolution of the emittance during propagation over 3000 time steps of the witness beam presented in Sec.~\ref{sec:correctness} with an initial transverse offset of the witness bunch centroid of $x_b = \sigma_x$, for the two field solvers, in normalized units. The error due to single precision is computed for each field solver with respect to the double-precision simulation; (b) Same for a simulation running in SI units, where $k_p^{-1}=10\,\mu$m.}
\label{fig:precision}
\end{figure}
The code can be compiled in either double (C++ \verb!double!) or single (C++ \verb!float!) precision, a feature inherited from AMReX. The effect of the precision on the simulation accuracy is investigated by comparing the evolution of the emittance of the witness beam of the reference setup with an initial emittance of $\epsilon_{x,0} = 0$ when an initial transverse offset of the bunch centroid $x_b = \sigma_x$ is present in Fig.~\ref{fig:precision}.
For both predictor-corrector and explicit field solvers, the error attributed to using single precision remains well below 1\% (2\%) in normalized units (SI units) after 3000 time steps. Table~\ref{tab:precision} shows the runtime in single and double precisions for the two solvers on two different architectures, a cutting-edge HPC GPU (NVIDIA A100) and a typical consumer-grade ("\emph{gaming}") GPU (NVIDIA RTX2070), easily available on a laptop. As anticipated, double-precision calculations are much faster on the HPC GPU than on the gaming GPU. However, with the capability to run high-resolution production simulations on a gaming GPU with comparable accuracy and performance as on an HPC GPU in single precision, HiPACE\texttt{++} provides useful scalability from laptops to the largest supercomputers.
\begin{table}
\begin{tabular}{|c|c||c|c|c|}
 \hline
GPU & Solver & $T_{double}$ & $T_{single}$ & speedup \\
 \hline
 \hline
A100 & P-C & 16.6 s & 12.9 s & 1.3$\times$ \\
A100 & explicit & 23.8 s & 20.6 s & 1.2$\times$ \\
 \hline
RTX2070 & P-C & 99.1 s & 36.3 s & 2.7$\times$ \\
RTX2070 & explicit & 66.6 s & 27.8 s & 2.4$\times$ \\
 \hline
\end{tabular}
\caption{Runtime for 1 time step of the simulation presented in Fig.~\ref{fig:precision} on NVIDIA A100 and NVIDIA RTX2070 GPUs. P-C stands for the predictor-corrector loop. \label{tab:precision}}
\end{table}

Though in active development, HiPACE\texttt{++} features numerous capabilities useful for production simulations including multiple beams and plasma with different species and profiles (driver and witness beam, ion motion, etc.), the possibilities to load external beams and apply external fields as well as specialized mesh refinement capabilities~\cite{Mehrling:AAC:2019}. Field ionization using the ADK-model \cite{Ammosov:1986} is available in SI units and could readily be extended to normalized units. HiPACE\texttt{++} proposes different field solvers, and uses modern software practices to make it a user-friendly and stable code (continuous integration, open development repository, extensive documentation and comments). Planned upgrades include a laser envelope model~\cite{Benedetti:2017}, full mesh refinement, and the support of more GPU architectures from other providers. The code is fully operational on CPU and on modern NVIDIA GPUs, and simplified simulations are already possible on modern AMD GPUs.

\section{Conclusion}

This paper presented the open-source, performance-portable, 3D quasi-static PIC code HiPACE\texttt{++}. The main adjustments required to port the quasi-static PIC loop to accelerated computing consist in (i) ensuring that all operations, including the beam operations, are performed within the loop over slices so little data needs to be stored on device memory and (ii) proposing a different longitudinal parallelization (pipeline) with which the amount of data communicated per rank scales down with the number of ranks, enabling excellent scalability for production simulations up to hundreds of GPUs.

Focusing on runtime rather than scalability, HiPACE\texttt{++} is not MPI-parallel transversally: each slice is computed on a single GPU, enabling orders-of-magnitude speedup with respect to CPU implementations. Transverse parallelization will be considered if it provides significant speedup without impacting the code clarity. Benchmarks show excellent agreement with legacy code HiPACE and full electromagnetic PIC code WarpX.

HiPACE\texttt{++} is built on top of cutting-edge libraries (in particular AMReX and openPMD-api) to harness top performance-portability and encourage open science, while improving sustainability. It enables production simulations of plasma acceleration from laptops to supercomputers.

\section*{Acknowledgements}
The authors gratefully acknowledge helpful discussions with T.J.~Mehrling, C.B.~Schroeder, J.~Osterhoff, T.~Wetzel, and B.~Diederichs. We gratefully acknowledge the Gauss Centre for Supercomputing e.V. (www.gauss-centre.eu) for funding this project by providing computing time through the John von Neumann Institute for Computing (NIC) on the GCS Supercomputer JUWELS Booster at J\"ulich Supercomputing Centre (JSC). We acknowledge the Funding by the Helmholtz Matter and Technologies Accelerator Research and Development Program. This research was also supported by the Exascale Computing Project (No. 17-SC-20-SC), a collaborative effort of the U.S. Department of Energy’s Office of Science and National Nuclear Security Administration. We acknowledge the support of the Director, Office of Science, Office of High Energy Physics, of the U.S. Department of Energy under Contract No. DE-AC02-05CH11231. This research used resources of the Oak Ridge Leadership Computing Facility at the Oak Ridge National Laboratory, which is supported by the Office of Science of the U.S. Department of Energy under Contract No. DE-AC05-00OR22725.

\appendix*

\section{Simulation parameters}
\label{sec:simparam}

The reference setup consists of a drive and a witness beam. Both beams are Gaussian with rms sizes $k_p \sigma_{\perp, d} = 0.3$, $k_p \sigma_{\zeta, d} = 1.41$ and $k_p \sigma_{\perp, w} = 0.1$, $k_p \sigma_{\zeta, w} = 0.2$. The peak densities are $n_{b, d}/n_0 = 10$ and $n_{b, w}/n_0 = 100$. The drive beam had an initial energy of $10\,$GeV and $0.1 \%$ rms energy spread, the witness beam had an initial energy of $1\,$GeV and no initial energy spread. The drive beam is located at the origin, the witness beam is centered around longitudinal position $k_p \zeta_{0, witness} = -5$. The beams are initialized at waist, either with a fixed number of particles per cell with a variable weight or with random positions and fixed weights. The beams initialized by a variable weight use 1 particle per cell. The randomly initialized beams use $10^6$ fixed weight particles per beam. For all simulations, the domain is, in units of $k_p^{-1}$, ($-8,8$), ($-8,8$), and ($-7,5$) in $x$, $y$, and $\zeta$. The varying simulation parameters for the presented studies are listed in Tab.~\ref{tab:simparam}. The time step in all simulations is $dt = 6\,\omega_p^{-1}$. 
\newline

\begin{table}[h]
\begin{tabular}{|r|r|r|r|r|r|}
\hline
\multicolumn{1}{|l|}{Fig.} & \multicolumn{1}{l|}{Solver}                            & \multicolumn{1}{l|}{$n_{x,y} \times n_\zeta $}                                    & \multicolumn{1}{l|}{$n_{ppc}$}                & \multicolumn{1}{l|}{beam}                                & \multicolumn{1}{l|}{$n_{steps}$}                    \\ \hline
2                          & \begin{tabular}[c]{@{}r@{}}P-C (30)\\ explicit\end{tabular} &\begin{tabular}[c]{@{}r@{}}$1024  \times 1024$\\ $1024  \times 1024$\end{tabular} & \begin{tabular}[c]{@{}r@{}}4\\ 4\end{tabular} & \begin{tabular}[c]{@{}r@{}}fixed ppc\\ fixed ppc\end{tabular} & \begin{tabular}[c]{@{}r@{}}1\\ 1\end{tabular}       \\ \hline
3                          & P-C (5)                                                    &  \begin{tabular}[c]{@{}r@{}}$512  \times 1024$\\ $2048 \times 1024$\end{tabular}   & \begin{tabular}[c]{@{}r@{}}1\\ 1\end{tabular} & \begin{tabular}[c]{@{}r@{}}random\\ random\end{tabular}       & \begin{tabular}[c]{@{}r@{}}1\\ 1\end{tabular}       \\ \hline
4                          & \begin{tabular}[c]{@{}r@{}}P-C (5)\\ explicit\end{tabular} & \begin{tabular}[c]{@{}r@{}}$2^n \times 1024$\\ $2^n \times 1024$\end{tabular}     & \begin{tabular}[c]{@{}r@{}}1\\ 1\end{tabular} & \begin{tabular}[c]{@{}r@{}}random\\ random\end{tabular}       & \begin{tabular}[c]{@{}r@{}}1\\ 1\end{tabular}       \\ \hline
6                          & \begin{tabular}[c]{@{}r@{}}P-C (1)\\ P-C (1)\end{tabular}      & \begin{tabular}[c]{@{}r@{}}$1024  \times 1024$\\ $2048  \times 2048$\end{tabular} & \begin{tabular}[c]{@{}r@{}}4\\ 1\end{tabular} & \begin{tabular}[c]{@{}r@{}}random\\ random\end{tabular}       & \begin{tabular}[c]{@{}r@{}}1000\\ 2048\end{tabular} \\ \hline
7                          & P-C (5)                                                    & $1024  \times 1024$                                                               & 1                                             & random                                                        & 3000                                                \\ \hline
\end{tabular}
\caption{Simulation parameters of the presented studies in the respective figure. P-C stands for predictor-corrector solver, $n_{ppc}$ refers to the number of plasma particles per cell, the \emph{beam} column indicates whether the beam was initialized as a random beam with a fixed number of particles and a fixed weight or by a beam with fixed particles per cell and a variable weight, and $n_{steps}$ denotes the number of time steps. The maximum number of iterations in the predictor-corrector solver is given between brackets in the Solver column.}
\label{tab:simparam}
\end{table}

For the HiPACE\texttt{++} and WarpX simulations on the JUWELS Booster, we used GCC 9.3.0, CUDA 11.0, OpenMPI 4.1.0rc1, CMake 3.18.0, and FFTW 3.3.8, except for the strong scaling using the temporal domain decomposition on the JUWELS Booster, which used GCC 10.3.0, CUDA 11.3, and OpenMPI 4.1.1. The GPU runs were compiled with nvcc 11.0.221 using the following flags: \texttt{-O3 -gencode=arch=compute\_80,code=sm\_80 -gencode=arch=compute\_80,code=compute\_80 -maxrregcount=255 --use\_fast\_math}.

The HiPACE\texttt{++} CPU runs were compiled using the following flags: \texttt{-O3 -DNDEBUG  -pthread -fopenmp -Werror=return-type}.

The legacy code HiPACE was compiled with ICC 19.1.2.254 using the flags: \texttt{-std=c99 -march=native -O3 -Os}.

On the laptop, we used GCC 8.4.0, CUDA 11.0, MPI 3.1, and CMake 3.20.3. The code was compiled with nvcc 11.0.194 using the following flags: \texttt{-O3 -DNDEBUG  --expt-relaxed-constexpr --expt-extended-lambda -Xcudafe --diag\_suppress=esa\_on\_defaulted\_function\_ignored -maxrregcount=255 -Xcudafe --display\_error\_number --Wext-lambda-captures-this --use\_fast\_math -Xcompiler -pthread}

Throughout the studies, we used AMReX v21.05 to v21.09 and HiPACE\texttt{++} from commit 3f2f4e15a607 to v21.09, except for the simulation using spatial domain decomposition, which was conducted on commit 11523c24c0f7c73ce3fe8d3424ede54565f58d50.
\FloatBarrier

\typeout{} 
\bibliography{main}

\end{document}